\documentclass[a4paper]{article}

\newcommand{\xmax}{\ensuremath{X_{\rm max}}}
\newcommand{\lsim}{\mathrel{\hbox{\rlap{\lower.55ex \hbox{$\sim$}} \kern-.3em \raise.4ex \hbox{$<$}}}}
\newcommand{\gsim}{\mathrel{\hbox{\rlap{\lower.55ex \hbox{$\sim$}} \kern-.3em \raise.4ex \hbox{$>$}}}}

\usepackage{icrc2013}

\title{The Non-Imaging CHErenkov Array
(NICHE): A TA/TALE Extension to
Measure the Flux and Composition of
Very-High Energy Cosmic Rays}

\shorttitle{The Non-Imaging CHErenkov Array
(NICHE)}

\authors{
John Krizmanic$^{1}$,
Douglas Bergman$^{2}$,
\& Pierre Sokolsky$^2$
for the TA/TALE Collaboration.
}

\afiliations{
$^1$ Universities Space Research Association, Columbia, Maryland 21044 USA \\
$^2$ Dept. of Physics \& Astronomy, University of Utah, Salt Lake City, Utah 84112 USA \\
}

\email{jkrizmanic@usra.edu}

\abstract{Co-sited with TA/TALE, the Non-Imaging CHErenkov Array (NICHE) \cite{NICHEcr2012} will measure the flux and nuclear composition of
cosmic rays from below $10^{16}$ eV to $10^{18}$ eV in its initial deployment. Furthermore, the low-energy threshold can be
significantly decreased below the cosmic ray knee via counter redeployment or by including additional counters. NICHE
uses easily deployable detectors to measure the amplitude and time-spread of the air-shower Cherenkov signal to achieve
an event-by-event measurement of \xmax\ and energy, each with excellent resolution. NICHE will have sufficient area and
angular acceptance to have significant overlap with the TA/TALE detectors to allow for energy cross-calibration.
Simulated NICHE performance has shown that the array has the ability to distinguish between several different
composition models as well as measure the end of Galactic cosmic ray spectrum. 
}


\begin{document}
\maketitle

\section{Scientific Motivation}

Measurement of the changing nuclear composition of High-Energy Cosmic Rays (HECRs) at energies around the Knee, $\gsim 10^{15}$ eV, provides a unique tool for understanding the evolution of the high-energy end of the galactic CR spectrum while providing a firm foundation for understanding the composition and spectrum of extragalactic UHECRs. However, the current understanding is muddled
due to uncertainties inherent to the measurement techniques and/or dependence on hadronic Monte Carlo simulation models required to interpret the data \cite{Kampert2012}. 
This ambiguity makes it difficult to confirm the widely adopted hypothesis that the Knee is the result of a Peters cycle\cite{Peters1960} due to rigidity-dependent cutoffs of the various galactic CR nuclear components.

One way to minimize this state of uncertainty is to perform a coordinated set of experiments to observe the composition of CRs over many orders-of-magnitude, using techniques that are directly sensitive to composition-dependent observables. One such sensitive observable for measuring CR composition in air showers is the depth of shower maximum, \xmax, which is measurable in both Fluorescence Telescopes (FTs) and Cherenkov Detectors (CDs). Historically, non-imaging CDs and FTs have not been used together to simultaneously measure air showers. However, by using both the lateral distribution of Cherenkov photons and the time-domain structure of their arrival times, the spacing of Cherenkov counters can be sufficiently increased to lead to a large enough aperture to overlap with lower-energy FTs such as TALE.
Thus, the measurement of the Cherenkov air shower signal, calibrated with air fluorescence measurements, offers a methodology to provide an accurate measurement of the nuclear composition evolution over a large energy range.
NICHE will use an array of widely-spaced, easily deployable, non-imaging Cherenkov detectors to measure the amplitude and time-spread of the air shower Cherenkov signal to extract CR nuclear composition measurements and to cross-calibrate the Cherenkov energy and composition measurements with that from TA/TALE fluorescence and surface detectors. 

\vspace{-2mm}
\section{The Non-imaging Cherenkov Technique}

Energetic electrons generated in an extended air shower (EAS) produce Cherenkov radiation if they move faster than the speed-of-light in the local medium. The index of refraction in the atmosphere increases with depth, yielding Cherenkov cones from altitudes 8-20 km to overlap in a ring of radius 120-140 m at the ground. Since the interior of the ring is filled by the portion of the air shower at lowest altitudes, showers that develop deeper in the atmosphere will have a larger interior-to-ring ratio. Electron transverse momentum smears the Cherenkov ring on the ground, but the inside-to-outside ratio remains composition, i.e. \xmax, dependent.

\begin{figure}
\centering
  \includegraphics[width=.45\textwidth]{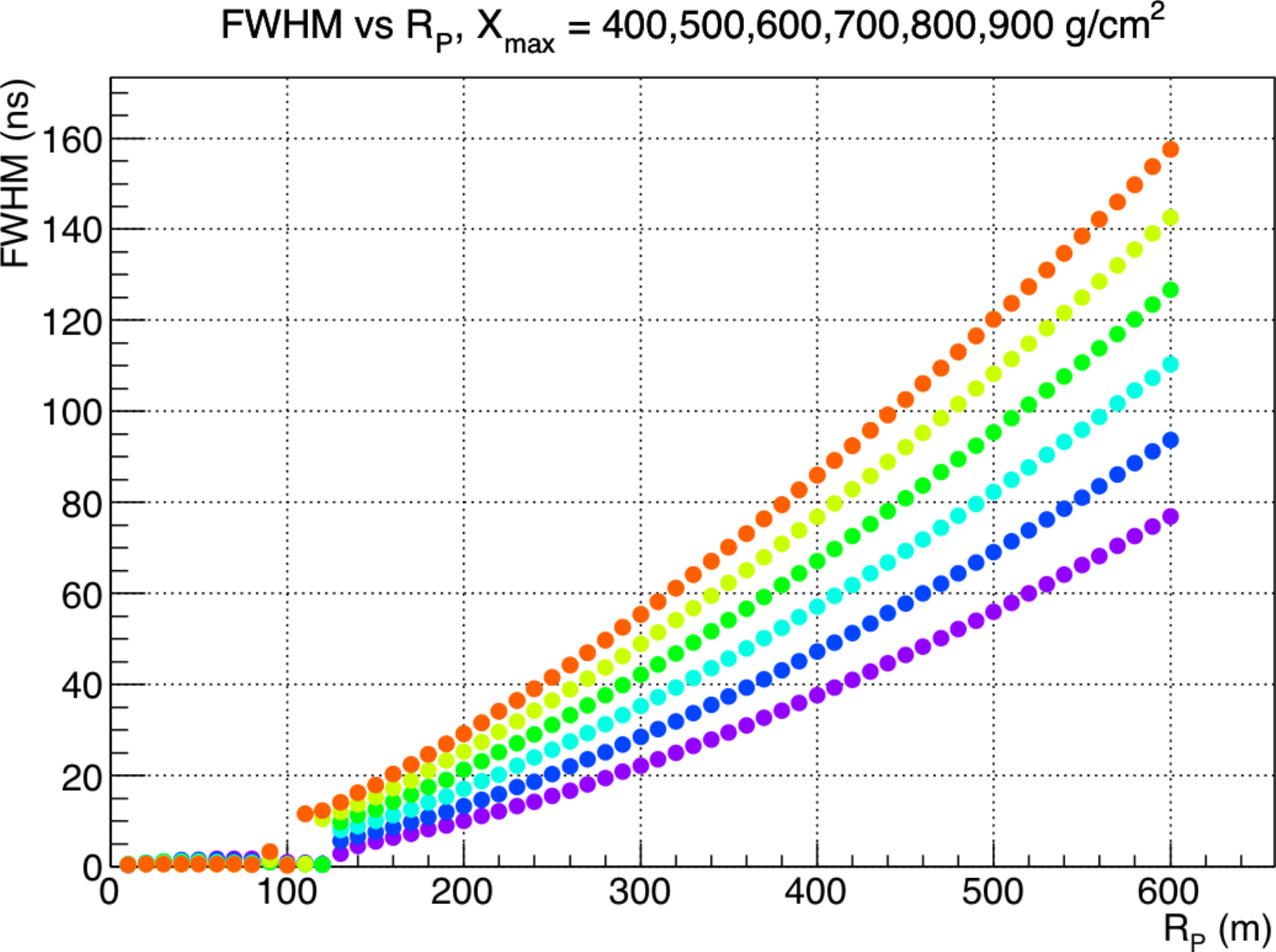}
  \caption{The simulated FWHM of the arrival time distribution of Cherenkov photons on the ground as a function of distance to the shower core for different \xmax\,  (vertical showers, $10^{16.5}$ eV). From low to high, \xmax= 400 g/cm$^2$ to 900 g/cm$^2$ in steps of 100 g/cm$^2$. The FWHM is discontinuous at the Cherenkov ring near 120 m, below which the FWHM significantly narrows.}
  \label{ctimewidth_fig}
\vspace{-2mm}
\end{figure}

At a given point on the ground, a counter will observe two components of Cherenkov light: one from the bulk of the shower, where some fraction of the transverse electrons are pointing their Cherenkov cones at the counter; and another part due to the smaller portion of the shower where the shower core Cherenkov cone intercepts the counter. Photons in the former component arrive over a long time span because the measurement samples a large portion of the developing air shower, while the latter component is narrow in time. When there is no dominant core component observed in a counter, the FWHM can be quite wide. Thus the FWHM in time provides a measure of shower development: deeper showers will have more Cherenkov light coming late due to sampling the air shower over a long path length. Figure \ref{ctimewidth_fig} illustrates how the temporal width for $10^{16.5}$ eV vertical showers varies as a function of \xmax\ as well as distance from the shower core, demonstrating that showers with deeper \xmax\, have a wider temporal FWHM for a given distance.

A number of experiments have employed the non-imaging Cherenkov technique (AIROBICC, BLANCA, CACTI, and Tunka) using the Cherenkov Light Distribution (CLD) to measure the CR spectrum, while two experiments (Tunka \& BASJE) have employed the Cherenkov Time Domain technique (references are listed in \cite{NICHEcr2012}). The innovation of NICHE is to combine these two techniques to construct an array of sufficiently large area to have significant overlap with TA/TALE air fluorescence measurements for $E \gsim 10^{17}$ eV, leading to a cross-calibration of the FT and non-imaging CD measurements.

\vspace{-2mm}
\section{NICHE Design}

\begin{figure}[t]
\centering
  \includegraphics[width=.4\textwidth]{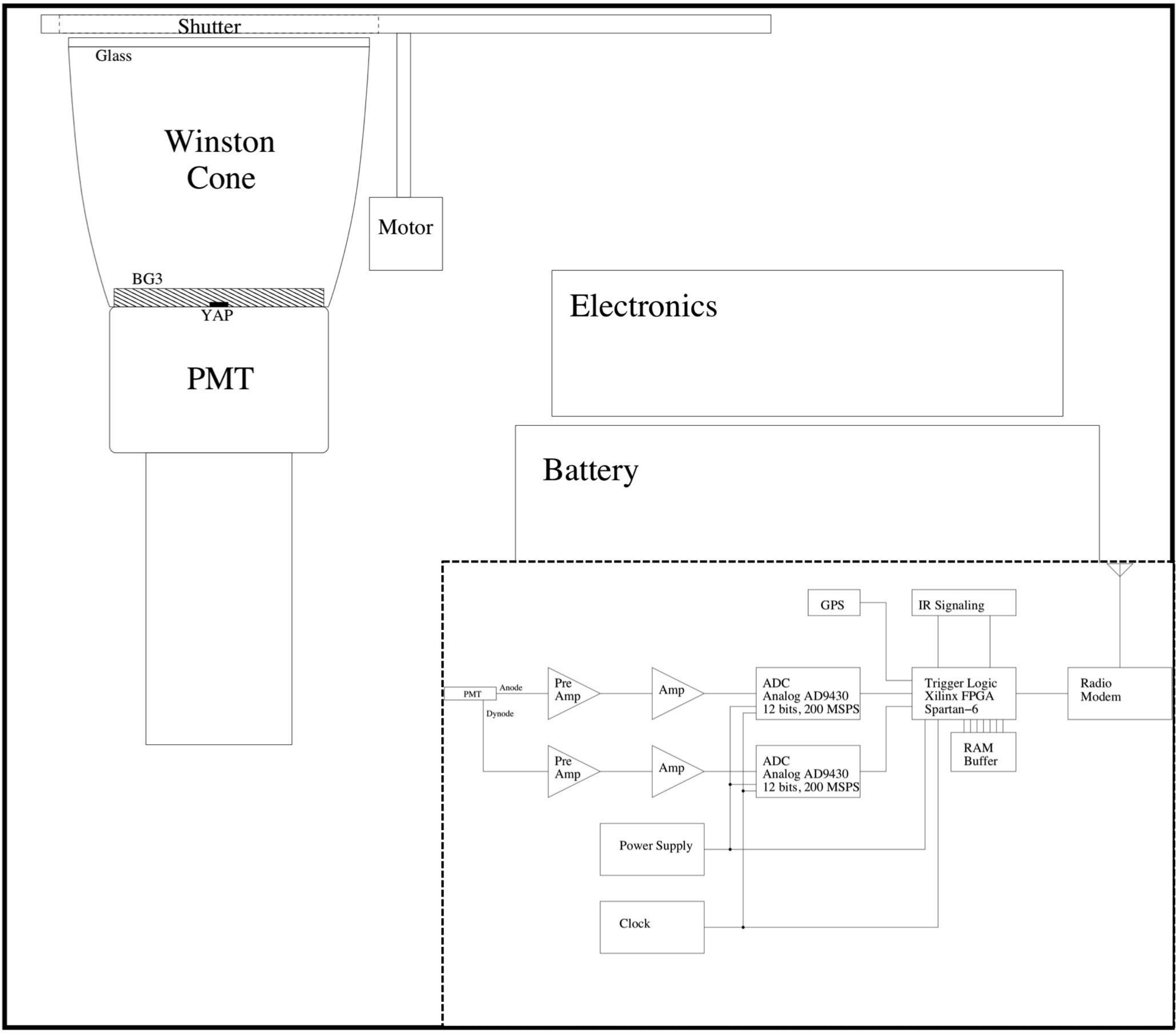}
  \caption{A schematic of the side view of single NICHE detector. The Inset is a block diagram of the data acquisition electronics for the detector.}
  \label{counter_fig}
\vspace{-2mm}
\end{figure}

NICHE's individual Cherenkov counters are based on the BLANCA counter design\cite{BLANCA2001}, and each counter uses a 3$^{\prime \prime}$ PMT with a Winston cone. Detailed CORSIKA-based simulation studies are being employed to optimize the design. The current baseline design uses a Winston cone 
with a 45$^\circ$ acceptance angle, leading to a 77 cm$^2$ active area for each counter. Key NICHE hardware improvements are to pair the PMT with a 200 MHz FADC DAQ system and to use GPS for signal time tagging. The counters are designed to be individually powered through battery and solar cells, and to communicate remotely through a radio-based WLAN system. Level-2 triggers will be provided by inter-counter IR communication. A schematic of a NICHE counter and its electronics is shown in the Figure \ref{counter_fig}.

\begin{figure}[b]
  \centering
  \includegraphics[width=0.47\textwidth]{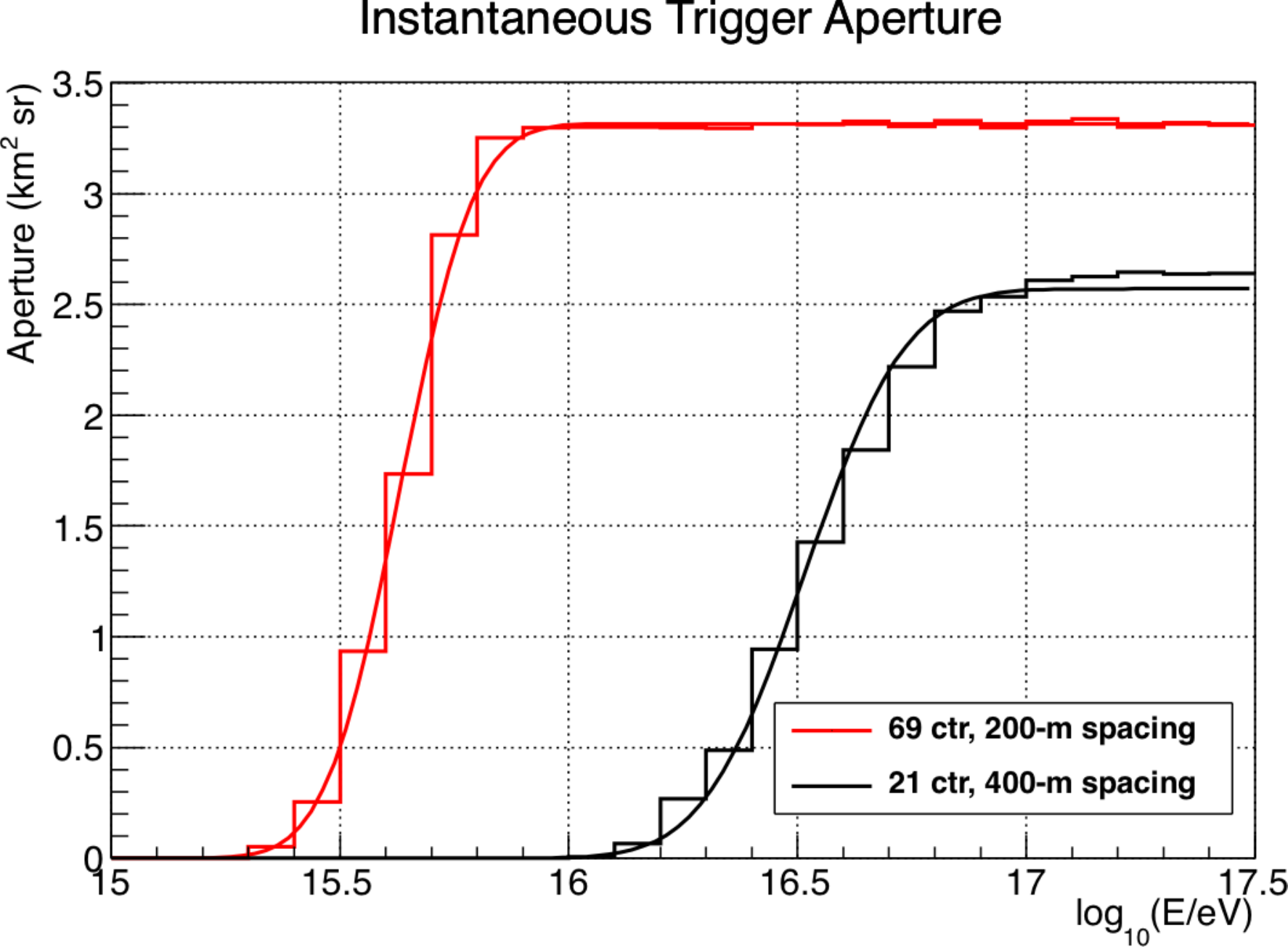}
  \caption{Simulated NICHE aperture as a function of energy.}
  \label{aperture_fig}
 \end{figure}

The NICHE array is designed to have enough aperture above $10^{17}$ eV to have a significant overlap with the TALE fluorescence and surface detector (SD) measurements.  In its initial deployment, the NICHE instrumented area will be nearly 2 km$^2$. Simulation studies indicate that a 69-counter array with 200-m counter spacing will have an instantaneous aperture of 3.3 km$^2$ sr above 10$^{15.8}$ eV assuming an angular acceptance up to zenith angles to 45$^\circ$. This corresponds to collecting $> 1500$ events per year above $10^{17}$ eV assuming a 10\% duty cycle. The NICHE energy-dependent baseline aperture is shown in Figure \ref{aperture_fig} compared to the aperture for a 21-counter array with 400-m spacing, which represents a step towards the 69-counter, 200-m spaced array.  The 200-m counter separation ensures that at least one counter, and usually more, provides measurements of the intensity and time-width of the Cherenkov light within the 120-m ring and with significantly more measurements outside the ring.

\begin{figure}
  \centering
  \includegraphics[width=0.45\textwidth]{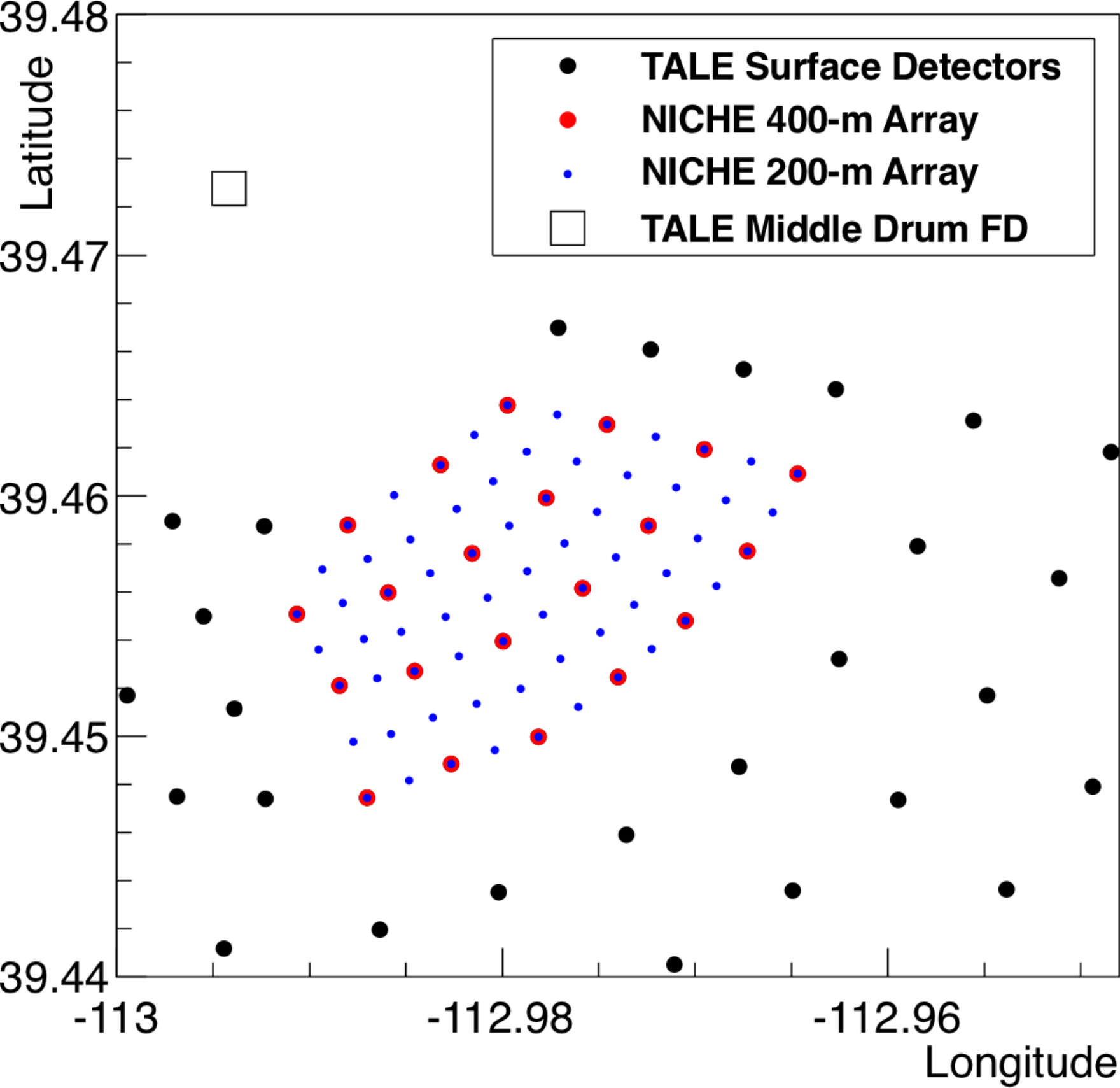}
  \caption{The expected layout of NICHE within the TA TALE extension with the Middle Drum FT station denoted by the square. The TALE SD detectors are large, black circles with the location of co-located NICHE detectors denoted by large, red circles. The small circles denote NICHE detectors that complete the 200-m spaced counter array.}
  \label{arrayloc_fig}
\vspace{-2mm}
 \end{figure}

The NICHE array will be located within the field-of-view of the TALE FT, as shown in Figure \ref{arrayloc_fig}, and also be interspersed with the TALE SDs. The NICHE array must be close to the FT detector to provide the largest possible overlap, in terms of energy, between the two disparate systems. The TALE detectors could add considerably to the aperture for time-domain measurements of NICHE as they will provide independent shower core position measurements for air showers that land outside of NICHE.
Eventually, a smaller infill array can be added to improve detection and reconstruction at low energies, where NICHE would act as a stand-alone detector. Low-energy events will trigger counters within a smaller radius, thus requiring the smaller counter spacing. The smaller aperture is compensated by a much larger CR flux at these lower energies.

\vspace{-2mm}
\section{NICHE Simulated Performance}

 The expected performance of a 200-m spaced, 69-counter array as well as a 400-m spaced, 21-counter array was determined from CORSIKA simulations. Samples of air showers for 5 different nuclear species were generated without thinning from 10$^{14}$ - 10$^{17.5}$ eV every 0.1 decade in energy with an $E^{-1}$ spectrum up to 45$^{\circ}$ in zenith angle. This event sample was then used to generate responses in a $41 \times 41$ array of counters with 50 m spacing and 77 cm$^2$ counter area.  A 1 pe/ns night-sky background was added, and the response of the detectors was modeled assuming a 9 ns PMT risetime followed by a 100 MHz, 2nd order Butterworth low-pass, anti-aliasing filter for the 200 MHz FADC response.  The level-1 trigger was defined by a 12-sigma short timescale fluctuation over a running average. A 5-fold counter coincidence was then applied to form a level-2 trigger.  Events were rejected if the largest counter signal was on the edge of the array to accept events that are fully contained in the NICHE array.  Note that the aperture could be increased by using the shower core measurements from the TALE detectors, and this is currently under study.

\begin{figure}[t]
  \centering
  \includegraphics[width=0.47\textwidth]{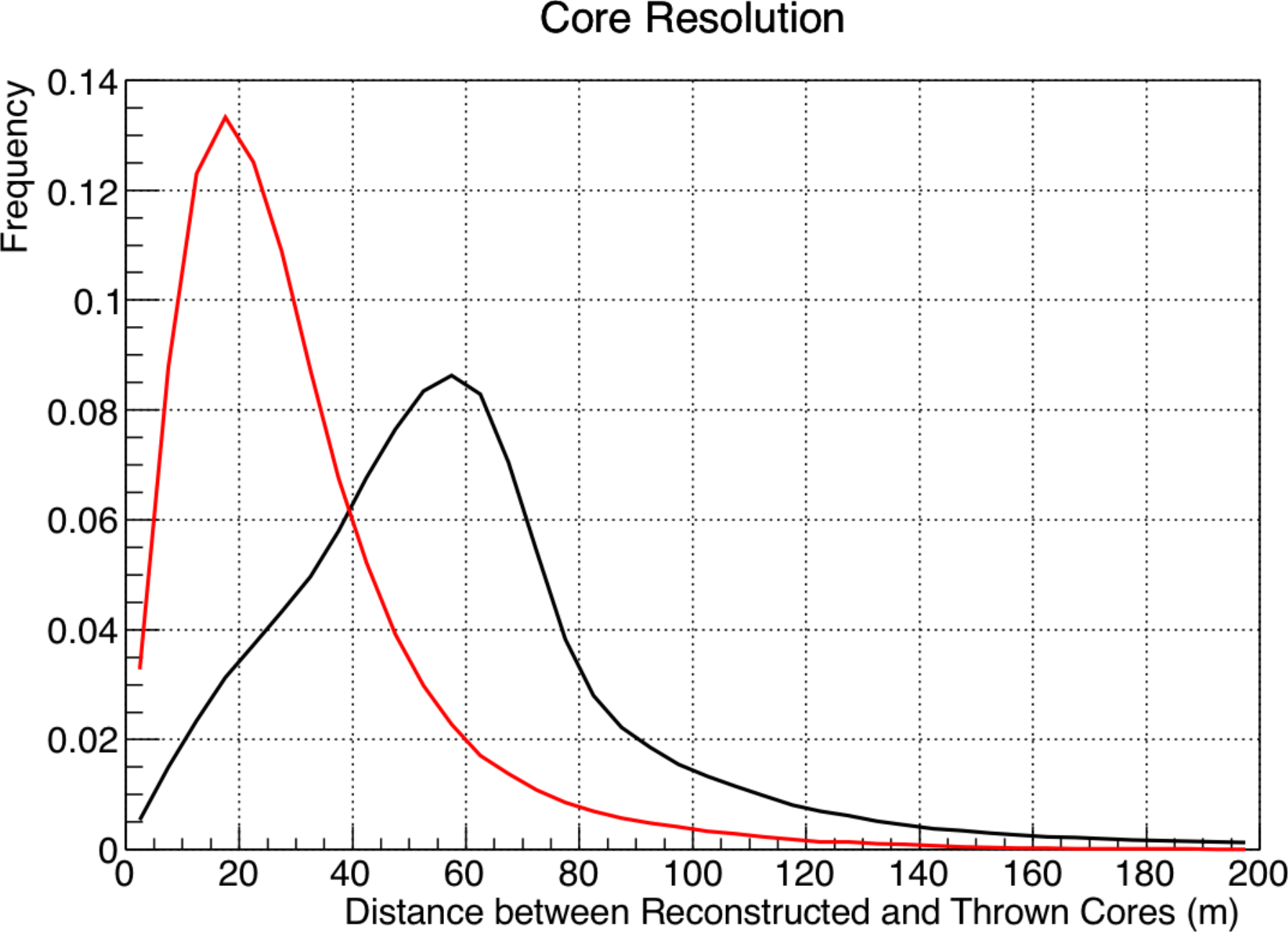}
  \caption{Simulated NICHE shower core resolution.}
  \label{coreres_fig}
\vspace{-2mm}
 \end{figure}

\begin{figure}[b]
  \centering
  \includegraphics[width=0.47\textwidth]{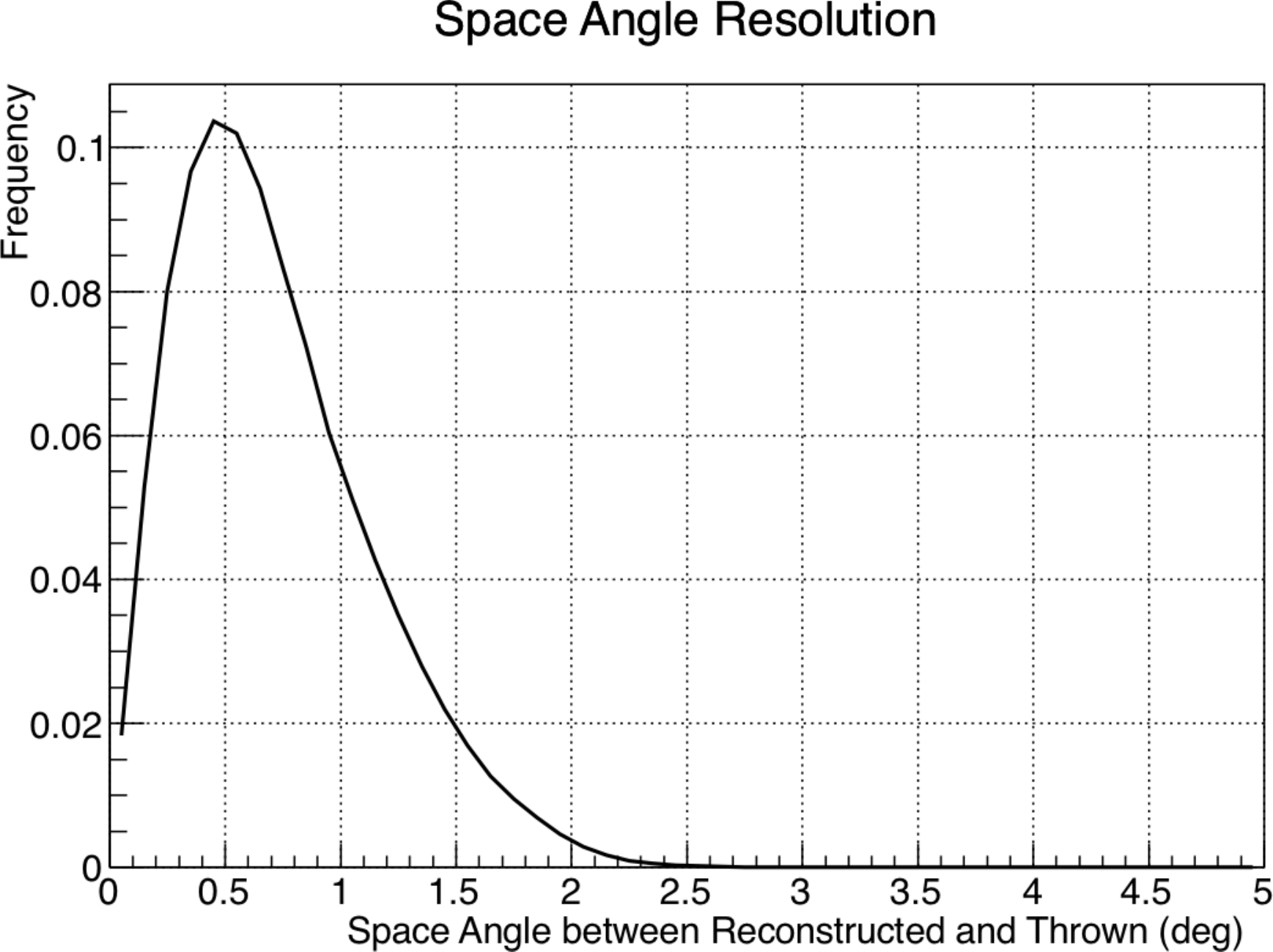}
  \caption{Simulated NICHE angular resolution.}
  \label{angres_fig}
 \end{figure}

Using this prescription as a guide, the response of arrays of Cherenkov detectors of various spacing, e.g. 50 m, 100 m, 200 m, 400 m, etc, can be modeled.  In this paper, arrays with 200-m and 400-m spacing are considered. For a fixed array spacing, the observed intensity and time profile of the Cherenkov signals in triggered detectors can be used to reconstruct the event.  An algorithm is employed to find the barycentric core of the event by using the cylindrical symmetry of footprint of triggered counters.  The frontside of the signal pulses (peaktime-risetime/2) is then used to determine the spatial angle of the event. The shower core position is then refined by using a functional fit of the time widths of the triggered counters.  This reconstruction procedure yields a core position resolution of $\sim 20$ m, shown in Figure \ref{coreres_fig}, and an angular resolution of $\sim 0.5^\circ$, shown in Figure \ref{angres_fig}.

An empirically determined formula is then used to reconstruct the energy and \xmax\ of the event for each triggered counter, based upon the knowledge of the distance from a particular counter to the shower core position on the ground.  The final event energy and \xmax\ are obtained by averaging the results from all triggered counters in the event. The simulated NICHE performance as a function of energy is then determined.

 \begin{figure}[t]
  \centering
  \includegraphics[width=0.47\textwidth]{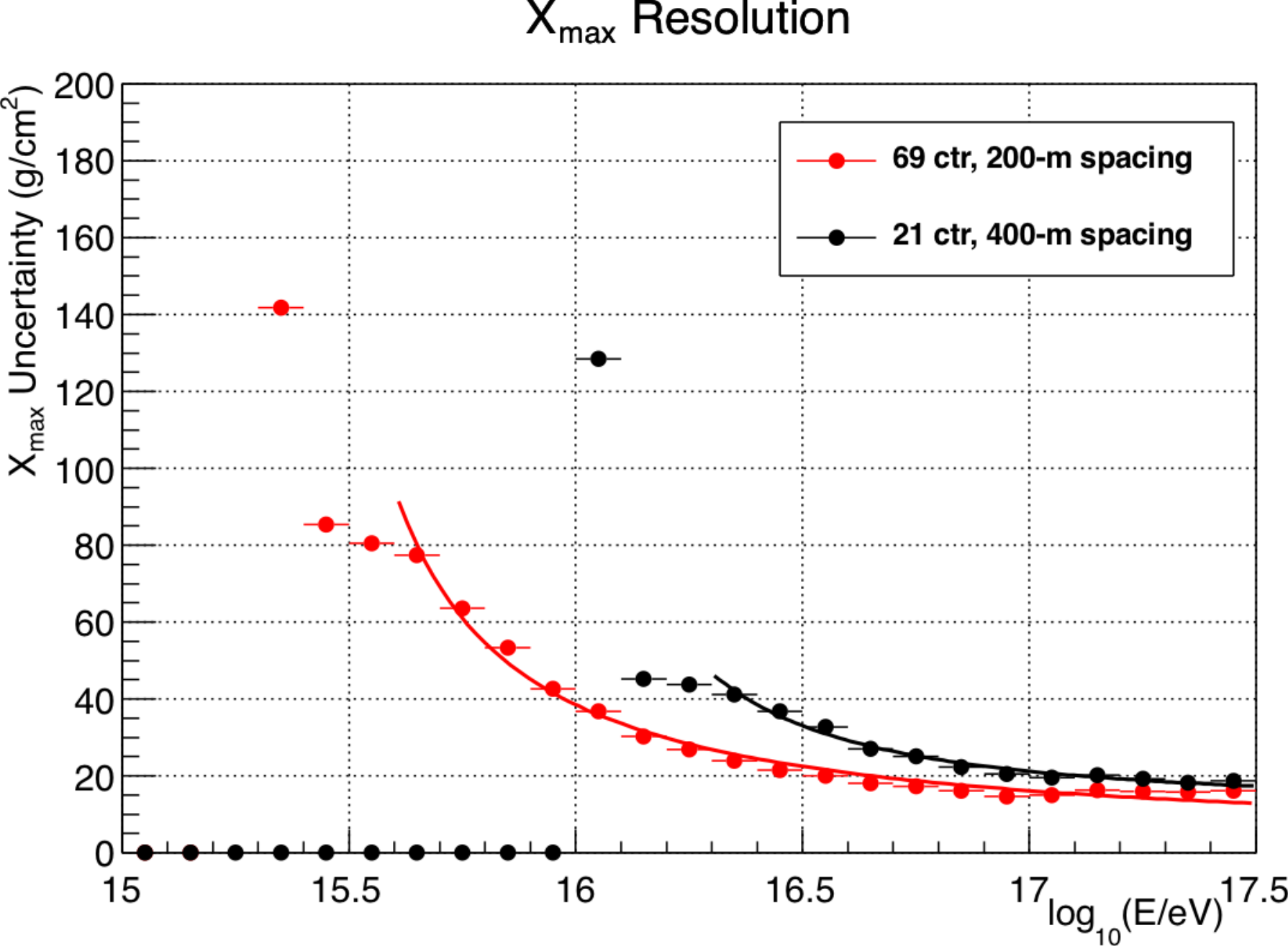}
  \caption{Simulated NICHE \xmax resolution as a function of energy.}
  \label{xmax_fig}
\vspace{-2mm}
 \end{figure}

 \begin{figure}[b]
  \centering
  \includegraphics[width=0.47\textwidth]{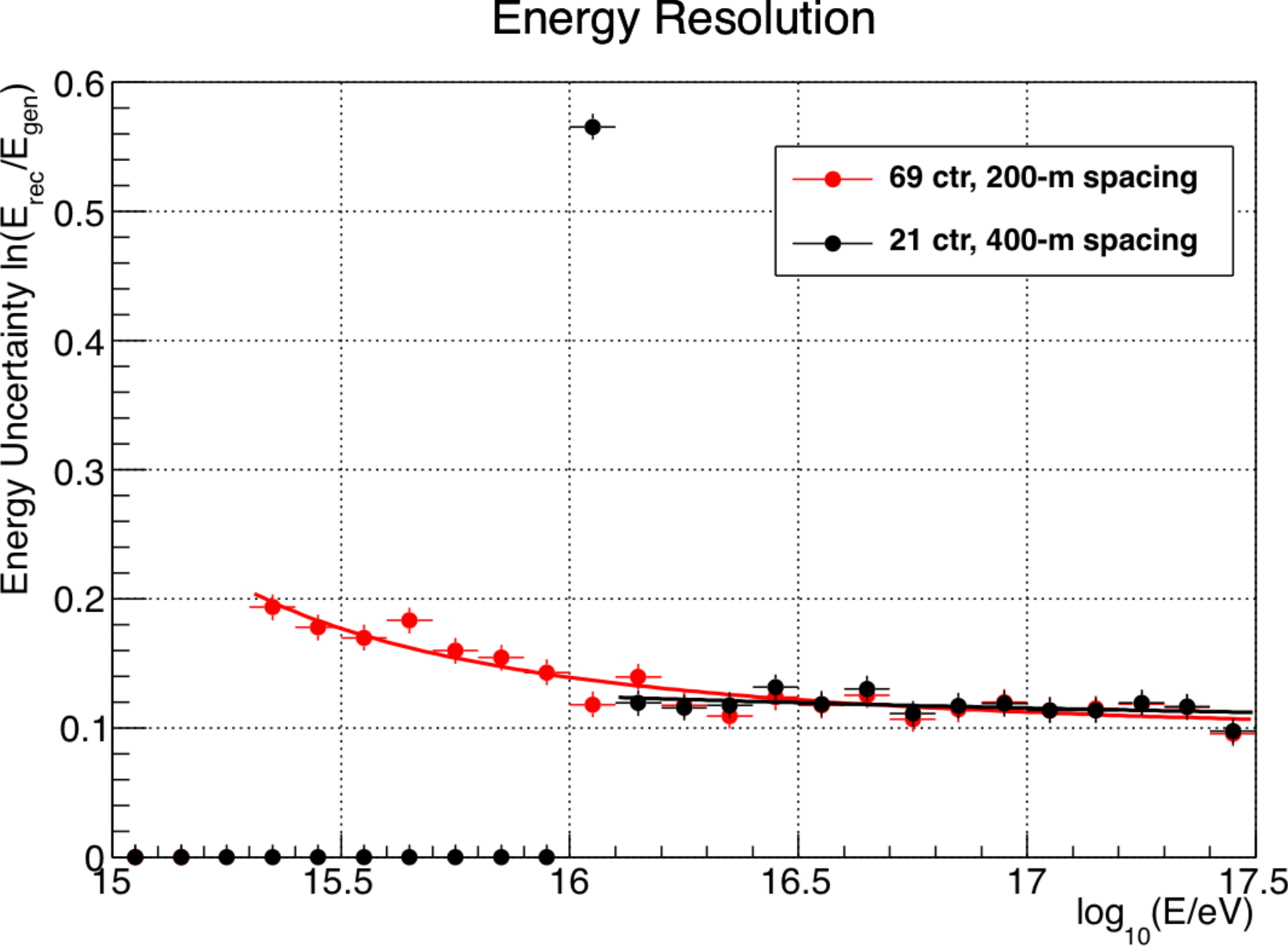}
  \caption{Simulated NICHE energy resolution as a function of energy.}
  \label{eres_fig}
 \end{figure}

Figure \ref{xmax_fig} shows the \xmax\ resolution for two NICHE configurations, a 69-counter array with 200-m spacing and a 21-counter array with 400-m spacing.  For the former, the \xmax\ resolution is less than 40 g/cm$^2$ at $10^{16}$ eV and decreases to better than 20 g/cm$^2$ above $10^{16.5}$ eV, while the \xmax\  resolution is modestly worse for the larger counter separation case. The energy resolution is presented in Figure \ref{eres_fig} and is better than 15\% at $10^{16}$ eV for the 200-m counter separation case and improves with increasing energy.

\vspace{-2mm}
\section{NICHE CR Composition Determination}

In order to quantify NICHE's ability to determine CR nuclear composition as a function of energy and to distinguish between different CR composition models, a procedure was developed {\cite{NICHEcomp} that employs high-statistics Monte Carlo generated distributions for a variety of individual CR nuclear components as a function of energy. The results were then used to construct simulated NICHE composition measurements, based upon poly-gonato CR composition models \cite{polygonato} and NICHE's simulated energy-dependent response. An unfolding procedure was then employed, using NICHE's simulated measurement performance for each specific nuclear species in the composition models, to determine the measured CR composition fractions as a function of energy for a the simulated composite spectrum.  

Specifically, a 4-component composition model was constructed using the energy-dependent nuclear composition ratios predicted by poly-gonato models. A 1-dimension air shower simulation \cite{Mikulski99} was used to generate high-statistics samples of \xmax\ distributions for proton, helium, CNO, and iron primary air showers. Composite spectra were then formed with the statistics determined for each third decade of energy based assuming two years of NICHE operation and 10\% duty cycle. The effects of \xmax\ and energy resolution were determined by the NICHE performance studies, based upon parameterizations as a function of energy of the results presented in Figures \ref{xmax_fig} and \ref{eres_fig}.  100 different independent runs were performed for each energy bin and included the effects of statistical fluctuations and variations in the \xmax\ for the events forming each sample.

Figure \ref{Ncomp_fig} presents the results of the modeling of NICHE's CR composition determination, in terms of $<\ln{A}>$, for two different models for the case of 69 detectors forming a 200-m spaced array overlaid on a sample of experimental measurements. One poly-gonato model has no extra-galactic component while the other has an extra-galactic proton component that leads to the total CR spectrum to be a pure power-law, assuming $\phi(E)=3 \times 10^{24} E^{-3}$ particles/(m$^2$ s sr eV) for $E \gsim 10^{16}$ eV.  While these two models are rather extreme in their composition dependence, they do highlight the ability for NICHE to distinguish between them at $10^{17}$ eV.  This is due to the excellent \xmax\ and energy resolution offered by the non-imaging Cherenkov technique as will be realized by NICHE. As discussed in \cite{NICHEcomp}, the unfolding of NICHE's simulated 4-component composition measurement reconstructs the thrown energy-dependent p, He, CNO, and Fe fractions to $\lsim 20$\% for most of the measurements, with larger errors on some points with low ($\lsim 5$\%) thrown fractions in the relevant composite spectrum.

 \begin{figure}
  \centering
  \includegraphics[width=0.47\textwidth]{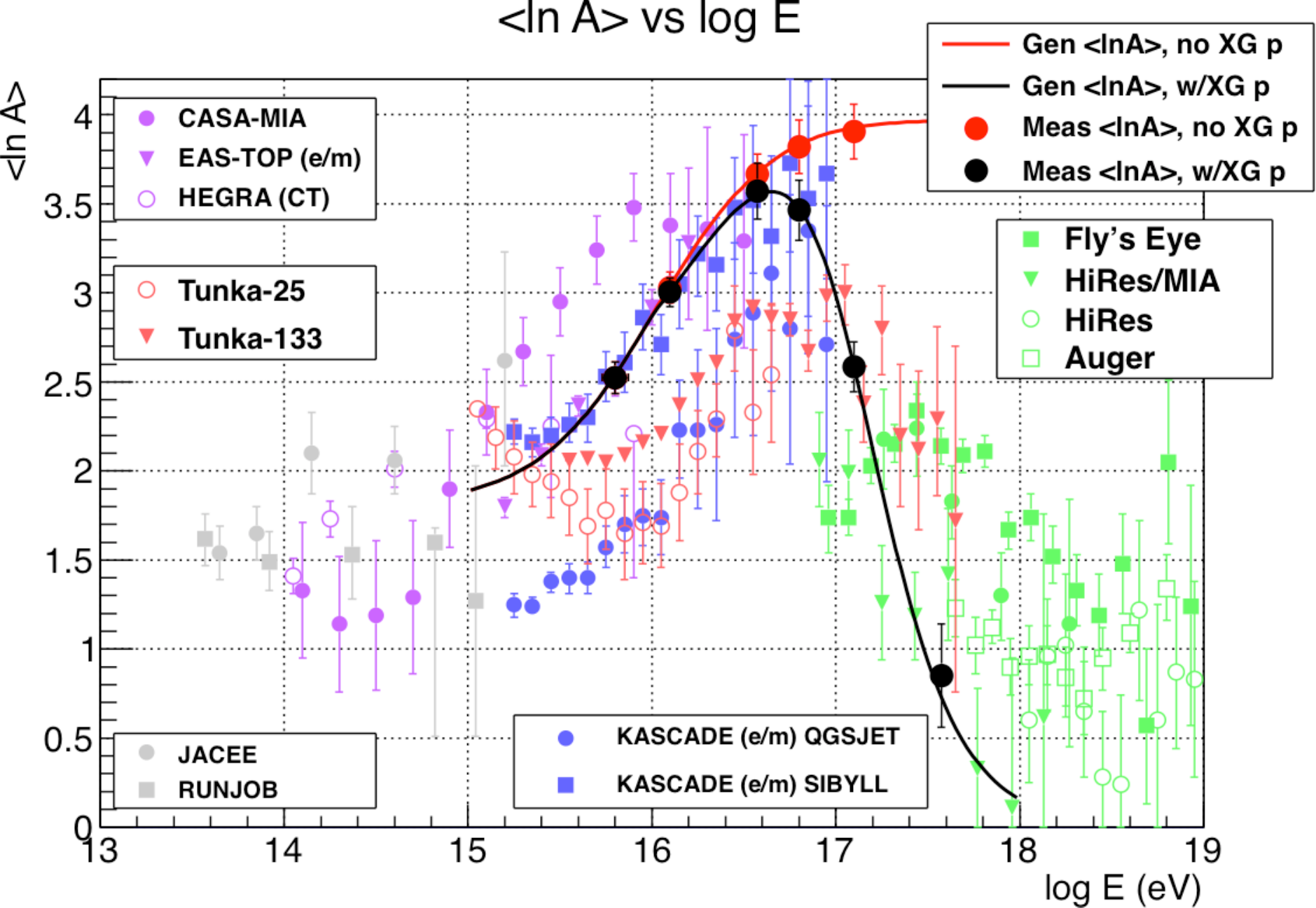}
  \caption{Comparison of NICHE nuclear composition performance compared to to current data (adapted from \cite{Blumer2009}).}
  \label{Ncomp_fig}
\vspace{1.5mm}
 \end{figure}

\vspace{-2mm}
\section{Conclusions}

NICHE will employ the non-imaging Cherenkov technique using an array on easily deployable detectors to perform cosmic ray flux and composition measurements for $E \gsim 10^{15.8}$ eV in NICHE's initial deployment.  Simulation studies indicate that NICHE will have a large statistics of events above $10^{17}$ eV that will allow for a cross-calibration with TALE's fluorescence measurements, assuming 2 year operation.  Simulation studies also indicate that the \xmax\  and energy resolution performance of NICHE will allow for the determination of at least a 4-component model of CR nuclear composition from $10^{15.8}$ up to $10^{18}$ eV, i.e. in the region where the galactic CR spectrum is hypothesized to be overtaken by an extra-galactic proton-like component.

\end{document}